\documentstyle[seceq,twoside,epsf]{ptptex}
\notypesetlogo  
\title{$S$-matrix Bases and Relation between\\
$\pi\pi$-Scattering and Production Amplitudes}
\author{%
Shin {\sc Ishida} and Muneyuki {\sc Ishida}$^{*}$ }
\inst{%
 Atomic Energy Research Institute, 
College of Science and Technology\\
Nihon University, Tokyo 101-0062, Japan\\
$^{*}$Department of Physics, Tokyo Institute of Technology\\
Tokyo 152-8551, Japan
}
\abst{%
It is stated that the requirement of unitarity and of analyticity should be made on the 
$S$-matrix elements with right bases, and the ``universality argument" made with the
bases, which do not regard quark physical picture, is not right. Accordingly
the result of the conventional analyses of the $\pi\pi$ production amplitudes 
following to this argument,
leading to non-existence of the light $\sigma$-meson, is proved to lose its 
theoretical bases. 
}

\begin{document}
\maketitle

\setcounter{tocdepth}{4}

\section{Purpose of This Talk}
The (partial $S$-wave) amplitude of $\pi\pi$-production ${\cal F}_{\pi\pi}(s)$ 
for $\sqrt s \stackrel{<}{\scriptstyle \sim} 1$GeV ($\sqrt s$ being total mass of the $\pi\pi$
system), obtained in most experimental processes such as $pp\to pp(\pi\pi )$, 
$\psi ' \to J/\psi (\pi\pi )$, $\Upsilon^{(n)}\to \Upsilon^{(m)} (\pi\pi )$ and $p\bar p\to (\pi\pi )\pi$
had been exceptionally crudely and  not duely treated\cite{rf1} 
under the influence of ``universality argument." 
It says\cite{rf2} that  ${\cal F}_{\pi\pi}(s)$ 
must be\footnote{
We treat, for simplicity, the case of single $\pi\pi$-channel in this talk. 
} 
proportional to 
the (partial $S$-wave) amplitude of $\pi\pi$-scattering  
${\cal T}_{\pi\pi}(s)$ 
as
\begin{eqnarray}
{\cal F} &=& \alpha (s) {\cal T},
\label{eq1}
\end{eqnarray}
where the $\alpha (s)$ is, due to the final state interaction (FSI) 
theorem,\footnote{
In this talk we assume that the theorem is valid, although it is, in the strongly interacting
system, not necessarily valid.
} real, and is moreover,
due to analyticity, slowly varying. Accordingly the structure of ${\cal F}$ obtained in any 
experiment should be the same as that of ${\cal T}$. In most of the conventional
works following to this 
argument, the analyses of ${\cal F}_{\pi\pi}(s)$ in this region 
$\sqrt s \stackrel{<}{\scriptstyle \sim} 1$GeV,
where the ${\cal T}(s)$ was investigated comparatively well, had become just fitting procedure
of experimentally obtained ${\cal F}_{\pi\pi}$ to ${\cal T}_{\pi\pi}$ through a respective function
$\alpha (s)$ expressed in terms of physically meaningless arbitrary parameters.
Since there was,\footnote{
This is, now considered, due to missing the cancellation mechanism, which originates from 
chiral symmetry, between the contributions to the ${\cal T}$ from the $\sigma$ meson and
the repulsive background interaction. 
(See, M. Ishida\cite{rfmy3}.)  
} 
at that time, observed 
no structure corresponding directly to the resonance with 
relevant mass in ${\cal T}_{\pi\pi}$, 
the large concentration\footnote{
This large concentration is now considered 
to represent the effect of $\sigma$-meson 
production. (See, T. Tsuru.\cite{rftsuru3})
}   
of iso-scalar $S$-wave $2\pi$ events (with mass 400$\sim$800MeV) frequently observed in the
many production channels had been conventionally treated as a mere background.
Thus the $\sigma$-particle, which was\cite{rf4} 
anticipated both theoretically and phenomenologically
in many works, had been disappeared for almost 20 years in the list of PDG before its 1996 
edition. 

The purpose of present talk is to point out\cite{rf1} that the FSI theorem 
and the analyticity\footnote{
Here it should be noted that  the basic field to expand the $S$-matrix bases gives generally
a physical origin of the singularity of amplitudes (see, \S 3).
} requirement 
should be applied on the $S$-matrix element with the ``right bases" and 
that the effective $\alpha (s)$ derived in a field theoretical model
with right quark-physical bases 
is not slowly varying function, that is, the 
above universality argument which dismisses quark picture is not correct.

Thus the analyses following the argument is meaningless, and the ${\cal F}(s)$ should be, 
in principle, treated independently from ${\cal T}(s)$.
Actually we made analyses\cite{rftsuru3} on the various production processes along this line,
and obtained strong evidences of $\sigma$-meson production.

\section{Phenomenological Methods of Analyses of Amplitudes}
In the following we summarize both methods in conventional and our
analyses and compare with them.

\hspace*{-0.7cm}[Conventional Method]
\begin{eqnarray}
{\cal T}(s) &=& {\cal K}(s)/(1-i\rho {\cal K}(s)), 
\label{eq2}\\
{\cal F}(s) &=& {\cal P}(s)/(1-i\rho {\cal K}(s)),\ \ 
{\cal P}(s)\equiv \alpha (s){\cal K}(s). 
\label{eq3}
\end{eqnarray}
These forms in Eqs. (\ref{eq2}) and (\ref{eq3}) are generally derived from the requirement
of elastic unitarity and of FSI theorem. 
However, the concrete forms of ${\cal K}(s)$ and ${\cal P}(s)$ (or $\alpha (s)$)
are model-dependent.
In the conventional method these are expressed in terms of arbitrary parameters 
with no physical meanings. For example, in the original analysis\cite{rf2} 
they choose a form, with arbitrary parameters, $\alpha_n$'s, and with the 
experimentally fixed zero-point of ${\cal T}(s)$, $s_0^{\cal T}$, as
\begin{eqnarray}
\alpha (s) &=& \sum_n \alpha_n s^n/(s-s_0^{\cal T}) .
\label{eq333}
\end{eqnarray}

\hspace*{-0.7cm}[Our Method] \\
\underline{Interfering Amplitude (IA) Method}:\cite{rf5}
\begin{eqnarray}
{\cal T}(s) &:& S = S^{\rm R} S^{\rm BG},\ \ S^{\rm R} = S^\sigma S^f ;\ \ \ 
 S^{(r)}=e^{2i\delta^{(r)}} \Rightarrow 
  \ \delta =\delta^\sigma +\delta^f + \delta^{\rm BG}; \nonumber\\
& S^{(r)} & \equiv 1+2ia^{(r)},\ \ \ 
a^{(r)} \equiv \sqrt\rho {\cal T}^{(r)} \sqrt \rho ; \ \ \   
a = a^{\rm R} + a^{\rm BG} + 2i a^{\rm R}  a^{\rm BG},\nonumber\\
 a^{\rm R}  &=& a^\sigma + a^f + 
2ia^\sigma a^f ,\ \ \  
a^\sigma = a_{\rm BW}^\sigma  \equiv 
\frac{\sqrt s \Gamma_\sigma (s)}{\lambda_\sigma-s}\ \ {\rm etc.} 
\label{eq4}
\end{eqnarray}
\underline{Variant Mass and Width (VMW) Method}:
\begin{eqnarray}
{\cal F}(s) &:&  {\cal F}= \frac{r_\sigma e^{i\theta_\sigma}}{\lambda_\sigma -s} 
                + \frac{r_f e^{i\theta_f}}{\lambda_f -s}
 + r_{\rm BG} e^{i\theta_{\rm BG}},
\ \ \   \lambda_\sigma  \equiv  m_\sigma^2 - i\sqrt s \Gamma_\sigma 
(s)\ \ {\rm etc};\ \ \ \  
\label{eq5}
\end{eqnarray}
where all terms and contained parameters have respective direct physical 
meanings.
In Eqs. (\ref{eq4}) and (\ref{eq5}) we give the formulas 
in the relevant case with two resonances, $\sigma$ and $f_0(980)$.
Our ${\cal T}$ given in Eq.(\ref{eq4}) satisfies automatically 
the elastic unitarity, while Eq. (\ref{eq5}) is consistent 
with the $S$-matrix unitarity. 
The FSI theorem gives some constraints\footnote{
These $r_i$ and $\theta_i$ are generally  real functions of $s$. However, 
it is shown\cite{rf1} that they are almost constant in the two 
resonance-dominating case. 
} among the $r_i$ and $\theta_i$ 
contained in ${\cal F}$. 
Here, it is to be noted that our ${\cal T}$ and ${\cal F}$
given above are also rewritten\cite{rf1} into 
the general forms Eqs. (\ref{eq2}) and (\ref{eq3}).
(See the following sections.)

\section{Strong Interaction and Bases of $S$-matrix}
Before going into detailed discussions on the relevant problem, 
it is necessary to
consider about rather general and fundamental physical situation:\\
The strong interaction is ``residual interaction" of QCD among color-neutral bound states
$\bar \phi$ of quarks (antiquarks) $q(\bar q)$ and gluons $g$. 
\begin{eqnarray}
{\cal H}^{\rm str} (\bar \phi_i) &=& {\rm resid.\ int.\ among\ }\bar \phi_i '{\rm s} .
\end{eqnarray}  
The unitarity of $S$ matrix is guaranteed by hermiticity of interaction 
Hamiltonian,
\begin{eqnarray}
SS^\dagger &=& S^\dagger S =1 \longleftarrow  {\cal H}^{{\rm I}\dagger }={\cal H}^{\rm I}.
\end{eqnarray}
It should also be noted that the basic fields are 
stable bound states $\bar\phi_i$'s
and that the complete set of bases of $S$-matrix is given 
as a configuration space of these
multi-$\bar\phi_i$ states.
Here it is instructive to remember the old history of strong interaction of the $\pi N$ system.

\begin{table}
\caption{Bases of $S$-matrix --- Old History of 
Pictures on Strong Interaction of $\pi N$ System}
\begin{tabular}{c|cl|l}
\hline\hline
               & Chew-Low & Theory & After quark physics \\
\hline
Switch off &  &  &      \\
${\cal H}^{\rm I}$ & Basic fields & $\pi$, $N$
  &  $\bar\pi =(q\bar q)$,  $\bar N,\ \underline{\bar\Delta =(qqq)}$ ``zero" $\Gamma$\\
\hline
Switch on &  &  &       \\
${\cal H}^{\rm I}$  & Resonance & $\Delta$ $(N\pi +N\pi\pi )$
  &  $\pi_{\rm phys.}$, $N_{\rm phys.}$; $\underline{\Delta_{\rm phys.}}$ ``finite" $\Gamma$\\
\hline
    & Compl. Set &    &    \\
    & of $S$-bases  & $| \pi \rangle$, $| N \rangle$, $| \pi N \rangle$, $\cdots$
 & $|\bar\pi\rangle$, $|\bar N\rangle$,
$\underline{|\bar\Delta\rangle}$, $|\bar\pi \bar N\rangle$, 
$\underline{|\bar\pi \bar\Delta\rangle}$, $\cdots$    \\
\hline
\end{tabular}
\end{table}

\begin{table}
\caption{Present problem}
\begin{tabular}{l|l|l}
\hline\hline
               & Unitary Chiral approach &  Ours (L$\sigma$M, NJLM) \\
\hline
Basic fields & $\bar\pi$  & $\bar\pi$; $\bar\sigma$, $\bar f$=$(q\bar q)$ ``zero" $\Gamma$ \\
Resonance & $\sigma (\pi\pi )$, $[f(K\bar K)]$  
  &  $\ \ \  \sigma_{\rm phys.}$\ \ $f_{\rm phys.}$\ \  ``finite" $\Gamma$\\
Compl. set of $S$-bases & $| \pi \rangle$, $| \pi\pi \rangle$, $\cdots$
 &   $|\bar\pi\rangle ,\ | \bar\sigma\rangle$, $|\bar f\rangle ,\ | \bar\pi\bar\pi\rangle$ \\
\hline
\end{tabular}
\end{table}

As is summarized in Table I,  there are the two pictures; the one is of Chew-Low
theory and the other one is of quark physics. In the former the basic fields are only 
$\pi$ and $N$, while in the latter they include also the bare $\bar \Delta$ field with zero-width
as a three-quark stable bound-state $\bar\Delta$=$(qqq)$.
After switching on ${\cal H}_{\rm I}^{\rm str.}$, in the former the physical $\Delta$-particle
appears as a resonance of $\pi N$ system, while in the latter the 
$\bar \Delta$ becomes $\Delta_{\rm phys.}$ with finite width.   
These two pictures may be phenomenolgically consistent with each other 
in so far as concerned with interactions of the $\pi N$ system.
However, we recognize presently the latter as a true one from the general and fundamental
viewpoint. 

In the present problem of the $\pi\pi$ system with two resonant particles
$\sigma_{\rm phys.}(600)$ and $f_{0,{\rm phys.}}(980)$ the similar situation 
as is summarized in Table II may be valid.
That is, the two pictures, the unitary chiral approach and 
our quark-picture\footnote{
In our covariant classification scheme 
both $\pi$-nonet and $\sigma$-nonet
belong to the $q\bar q$ ``relativistic $S$-wave" states. 
(See S. Ishida.\cite{rfshin3})
} based on the 
linear $\sigma$ model (L$\sigma$M) and on the NJL model, may be phenomenologically
consistent. But we consider that the latter should be 
the very true one from the general and
fundamental viewpoint.

\section{Relation between Scattering and Production Amplitudes}
\hspace*{-0.7cm}[\underline{Field Theoretical Model}]\ \ \ In order to 
explain clearly 
the essential points of our problem we consider a simplified field theoretical model of the relevant
system, where 
we should take the bare fields $\bar\sigma$ and $\bar f$
 as well as the $\bar\pi$ as basic fields, and 
we set up the strong interaction Hamiltonian
\begin{equation}
H_{\rm int}^{\rm scatt}=\sum_{\alpha =\sigma ,f}\bar g_\alpha\bar\alpha\pi\pi
+\bar g_{\pi\pi}(\pi\pi )^2,\ \ 
H_{\rm int}^{\rm prod}=\sum_{\alpha =\sigma ,f}\bar \xi_\alpha\bar\alpha ``P''
+\bar\xi_{\pi\pi}\pi\pi ``P'',
\label{eq7}
\end{equation}
where $\bar g$ and $\bar\xi$ are real coupling constants, and ``P'' denotes a 
relevant production channel. 
Taking into account the pion-loop effects due to the $H_{\rm int}^{\rm scatt}$, 
the stable bare states $\bar\pi ,\bar\sigma$ and $\bar f$ change into the
physical states denoted as $\pi =(\bar\pi )$, and $\sigma$ and $f$ with 
finite widths. Then we can derive the scattering and production amplitudes 
following the standard procedure of quantum field theory.

\begin{figure}[t]
 \epsfysize=4.5 cm
 \centerline{\epsffile{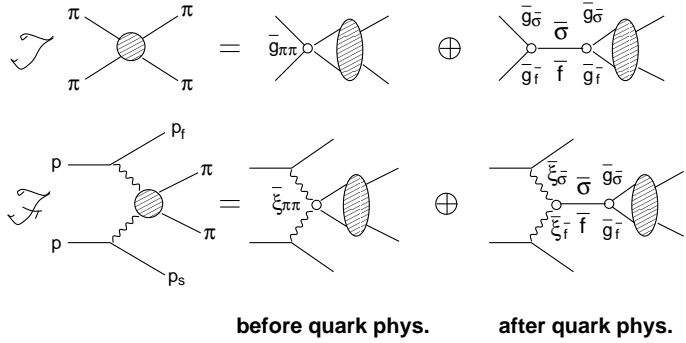}}
 \caption{The mechanism for scattering amplitude ${\cal T}$ and 
      production amplitude ${\cal F}$. The second diagram should be 
      correctly taken into account as well as the first, whereas
      only the first had been considered in the conventional treatment. 
    \label{figcomp} }
\end{figure}

The general structure of ${\cal T}$ and ${\cal F}$ is shown schematically in
Fig. \ref{figcomp}, where shaded ellipses represent the final state 
interaction of the $2\pi$ system. 
It is to be noted that correctly 
both the mechanisms in Fig. \ref{figcomp} should be taken into account. 
As a matter of fact, 
in the conventional treatment, where 
only the former is taken into account,
the $\alpha (s)$ in Eq.(\ref{eq3})
becomes
\begin{eqnarray}
\alpha (s) &=& \bar\xi_{\pi\pi}/\bar g_{\pi\pi}={\rm const}.\ ,
\end{eqnarray}
which is surely a (most) slowly varying function with $s$.

\begin{figure}[t]
 \epsfysize=2.5 cm
 \centerline{\epsffile{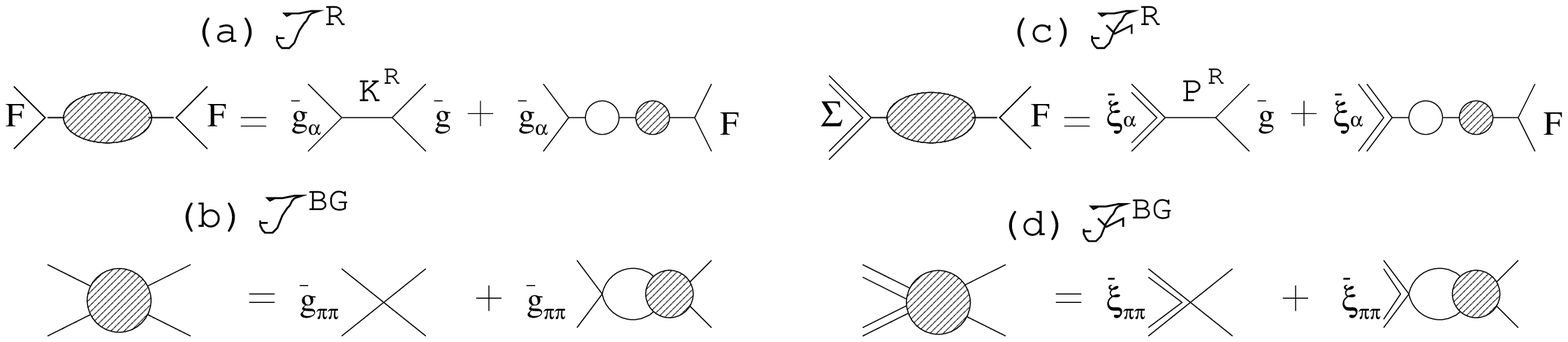}}
 \caption{
Scattering and production mechanism 
in a simple field-theoretical model.
The respective production amplitudes ${\cal F}^{(r)}$ are obtained, 
following the mechanism shown in the figures, by replacing the 
first $\pi\pi$-coupling constant $\bar g$ in ${\cal T}^{(r)}$
with the production coupling $\bar\xi$.  
The  ${\cal F}$ amplitude obtained in this 
way automatically satisfies the FSI theorem. 
 }
\label{fig:sd}
\end{figure}
In the previous work\cite{rf1} resorting to the above model 
we have derived our methods of
analyses, the IA method for ${\cal T}$ and the VMW method for ${\cal F}$, 
and shown their consistency with the FSI theorem.
The obtained formulas of the amplitudes
(derived as solutions of Schwinger-Dyson equations shown in Fig. 2
where (a) and (b) ( (c) and (d) ) correspond,\footnote{
Here we apply a rather special method of unitarization, which keeps the physical 
meanings of the respective tree diagrams such as resonances, 
backgrounds and so on.
} respectively, to ${\cal T}^R$ and ${\cal T}^{BG}$ 
(${\cal F}^R$ and ${\cal F}^{BG}$))
are given for quantities defined in the general formulas, 
Eqs. (\ref{eq2}), (\ref{eq3}) and (\ref{eq4}), as 
\begin{eqnarray}
{\cal T} &=& \frac{{\cal K}^R+{\cal K}^{BG}}{(1-i\rho{\cal K}^R)
(1-i\rho{\cal K}^{BG})} ,\ \ \  
{\cal K} = \frac{{\cal K}^R+{\cal K}^{BG}}{1-\rho^2 {\cal K}^R{\cal K}^{BG}},
\label{eq43a}\\
{\cal F} &=& \frac{{\cal P}^R+{\cal P}^{BG}}{(1-i\rho{\cal K}^R)
(1-i\rho{\cal K}^{BG})},\ 
{\cal P}^R = \frac{{\cal P}_\sigma +{\cal P}_f}
{1-\rho^2{\cal K}_\sigma{\cal K}_f},\ \ \ 
S^{(r)}\equiv \frac{1+i\rho {\cal K}^{(r)}}{1-i\rho {\cal K}^{(r)}}
 \label{eq44a}\\
{\cal T} &=& {\cal T}^R(1+i\rho{\cal T}^{BG})
+{\cal T}^{BG}(1+i\rho{\cal T}^R),\ \  
{\cal F} = {\cal F}^R(1+i\rho{\cal T}^{BG})
+{\cal F}^{BG}(1+i\rho{\cal T}^R) \nonumber\\
{\cal K}^R &=& \frac{{\cal K}_\sigma +{\cal K}_f}{
1-\rho^2 {\cal K}_\sigma{\cal K}_f},\ \ \ 
{\cal K}_\alpha  = \frac{\bar g_\alpha^2}{\bar m_\alpha^2 -s},
\ \ (\alpha =\sigma ,f),\ \ \ {\cal K}^{BG}=\bar g_{\pi\pi} \label{43b}\\
{\cal P}_\alpha &=& \frac{\bar\xi_\alpha \bar g_\alpha}{m_\alpha^2-s};
\ \ {\cal P}^{BG} = \bar\xi_{\pi\pi} , 
\ \ \  
S^R
=\Pi_{\alpha =\sigma ,f}\frac{\lambda_\alpha^*-s}{\lambda_\alpha-s},
\ \ \ 
S^{BG} = \frac{1+i\rho \bar g_{\pi\pi}}{1-i\rho \bar g_{\pi\pi}} ,\ \ \ \ \ \  
\label{eq12}
\end{eqnarray}
where the formulas given in Eqs. (\ref{eq43a}) and (\ref{eq44a})
are rather generally derived following 
the mechanism of Fig. 2, while those in Eqs.(\ref{43b}) and (\ref{eq12}) 
are derived depending on our choice of ${\cal H}_{\rm int}$ (\ref{eq7})
in the ``bare-state representation.'' 
These formulas of ${\cal T}$ and ${\cal F}$ are rewritten
into the forms of Eq.(\ref{eq4}) and Eq.(\ref{eq5}), respectively,
in the ``physical state representation.''\footnote{
Here we treat a simple case of including
only the virtual two-$\pi$ meson effects. In the above equations we also made 
simplification by identifying the ``${\cal K}$-matrix states''
(having diagonal (non-diagonal) real (imaginary) parts of mass matrix) with the 
bare states. As for details see Ref. \citen{rf1}
}
 The consistency with the FSI theorem of these 
amplitudes ${\cal F}$ and ${\cal T}$ is easily seen from 
Eqs.(\ref{eq43a}) and (\ref{eq44a}), 
since ${\cal K}$ and ${\cal P}$ are real and 
their phases come only from their common denominator 
$(1-i\rho {\cal K}^R)(1-i\rho {\cal K}^{BG})$. 
The above amplitudes lead to
\begin{eqnarray}
\alpha (s) &=& 
\frac{\bar \xi_\sigma \bar g_\sigma (m_f^2-s)+\bar \xi_f \bar g_f (m_\sigma^2-s)
+\bar \xi_{\pi\pi}((m_\sigma^2-s)(m_f^2-s)-\rho^2\bar g_\sigma^2\bar g_f^2)}
{\bar g_\sigma^2(m_f^2-s)+\bar g_f^2(m_\sigma^2-s)
+\bar g_{\pi\pi}((m_\sigma^2-s)(m_f^2-s)-\rho^2\bar g_\sigma^2\bar g_f^2)},
\ \ \ \ 
\end{eqnarray}
which is represented in terms of all physically meningful parameters.
This has a form in
contradistinction to the choice Eq.(\ref{eq333}),
and is not a slowly varying function.
In our method the large event concentration mentioned in \S 1 is directly
understood as results of $\sigma$-production by taking as
 $\bar\xi_\sigma / \bar g_\sigma \stackrel{>}{\scriptstyle \sim} \bar\xi_f / \bar g_f 
\gg \bar\xi_{\pi\pi} / \bar g_{\pi\pi} \approx 0$ . 

\hspace*{-0.7cm}[\underline{Phenomenological Analyses}]\ \ \ In the actual 
analyses of ${\cal T}$ and ${\cal F}$ 
we have applied 
the IA method Eq.(\ref{eq4}) and the VMW method Eq.(\ref{eq5}), respectively, 
and obtained the strong evidence for existence of the light $\sigma$-meson:
In the analyses of scattering amplitudes we have chosen the form of ${\cal T}_{BG}$
wih a hard core type $\delta_{BG}=-p_1 r_c$ ($p_1$ being the pion momentum
in the $2\pi$ rest frame). In the analyses of production amplitudes we have applied the
two kinds of ${\cal F}$, the one which regards the constraints from the conventional 
FSI theorem and the other\cite{rf6} 
which does not (, but is consistent with the $S$-matrix
unitarity with the right quark-physical bases). 
An example of the former is given as 
Eq.(\ref{eq44a}) with ${\cal K}^{BG}=-
({\rm tan}p_1r_c) / \rho$ and
${\cal P}^{BG}=( \bar \xi_{2\pi} / r_c) {\cal K}^{BG}$
instead of ${\cal K}^{BG}$ and ${\cal P}^{BG}$ 
given, respectively, in Eqs.(\ref{43b}) and (\ref{eq12}).


\section{Concluding Remarks}

We have explained why the most conventional methods of analyses on ${\cal F}$ 
($s \leq 1$GeV$^2$) following the universality argument are physically meaningless 
and reviewed a new method of analysis taking quark-physical viewpoint correctly
into account. As a result we may conclude; on experiment that production
experiments have their own value independent of scattering experiments;
on hadron phenomenology that its important task should be to extract the information
on intrinsic  quark-physical parameters such as $\bar g_\alpha$ and $\bar m_\alpha$;
and on hadron theory that to explain 
their values from the fundamental theory, QCD or else,
is one of its important purpose.


\begin{thebibliography}{9}
\bibitem{rf1} S. Ishida; M. Ishida, in proc. of WHS99, Frascati, 1999, ed. by 
T.Bressani, A. Feliciello and A. Filippi,
Frascati Physics Series XV (1999), 85; 115. 
M. Ishida, S. Ishida and T. Ishida, 
   Prog. Theor. Phys. {\bf 99} (1998), 1031.
\bibitem{rf2} M. R. Pennington, in proc. of Int. Conf. Hadron' 95, 
Manchester UK, {\it July\ 1995}, ed. by M. C. Birse, G.D. Lafferty and J.A.McGovern,
(World Scientific, Singapore), p.3. 
K. L. Au, D. Morgan and M. R. Pennington ,Phys. Rev. {\bf D35} (1987), 1633.
\bibitem{rfmy3}M. Ishida, in proc. of WHS99; this proceedings.
\bibitem{rftsuru3}T. Tsuru, this proceedings. K. Takamatsu, in proc. of Hadron97 at BNL, 1997,
ed. by ed. by S.U. Chung and H.J. Willutzki, AIP conf. proc. 432. 
\bibitem{rf4} T. Hatsuda and T. Kunihiro, Phys. Rep. {\bf 247} (1994), 221;
this proceedings.
\bibitem{rfshin3}S. Ishida, M. Ishida and T. Maeda, Prog. Theor. Phys. {\bf 104} (2000), No.4.
\bibitem{rf5} S. Ishida et al., Prog. Theor. Phys. {\bf 95} (1996), 745.
\bibitem{rf6} M. Ishida, under preparation. 
\end{thebibliography}
\end{document}